\documentclass[journal=ancac3,manuscript=article]{achemso}

\usepackage[T1]{fontenc}       
\usepackage{epstopdf}
\usepackage{verbatim}
\usepackage{epsfig}
\usepackage{graphicx}
\usepackage{color}
\usepackage{textcomp}
\usepackage{float}

\author{Aitzol Garcia-Etxarri}
\affiliation{Donostia International Physics Center (DIPC), Donostia - San Sebastian 20018, Spain}
\email{aitzolgarcia@dipc.org}

\title[ \textsf{ancac3}] {Polarization singularities on high index nanoparticles}
  
\begin{document}


%
%
%
%
%

\begin{abstract}
In this article, we study the emergence of polarization singularities in the scattered fields of optical resonators excited by linearly polarized plane waves. First, we prove analytically that combinations of isotropic electric and magnetic dipoles can sustain $L$ surfaces, and $C$ lines that propagate from the near-field to the far field. Moreover, based on these analytical results, we derive anomalous scattering Kerker conditions trough singular optics arguments. Secondly, through exact full-field calculations, we demonstrate that high refractive index spherical resonators present such topologically protected features. Furthermore, we calculate the polarization structure of light around the generated $C$ lines, unveiling a M\"{o}bius strip structure in the main axis of the polarization ellipse. These results prove that high-index nanoparticles are excellent candidates for the generation and control of polarization singularities and that they may lead to new platforms for the experimental study of the topology of light fields around optical antennas.  
\\
\end{abstract}




Mathematical singularities may be defined as points where the value of a function in not defined or is not well behaved. In physics, singularities arise whenever a mathematical formalism fails at describing some particular physical phenomenon. Often, these singularities drop hints on interesting phenomena happening in a particular physical process. Singularities in wave physics have been studied since 1830 in different kinds of waves 
(for a good overview, see \cite{Berry:2000km}). In wave optics, two general types of singularities have been identified and studied. \cite{Dennis:2001vi}

In certain situations, the polarization of light is uniform throughout space and thus, light can be mathematically described as a position dependent complex scalar wave multiplied by a constant polarization vector. In these situations, singularities arise whenever the amplitude of the wave is zero. In such cases, even though the amplitude of the wave si well defined, its phase cannot be unambiguously determined. These singularities receive the name of \textit{Phase singularities}. Around these singularities, the phase of the scalar field varies gradually from $0$ to $2\pi q$, where $q$ is an integer (positive or negative) named the topological charge of the singularity. These phase singularities, also known as nodal lines or vortices, have been widely studied in the recent past due to their ability to carry orbital angular momentum in their vicinity \cite{Soskin:1997wj,MolinaTerriza:2007ig}. 





Nevertheless, in most situations, light presents a spatially varying polarization structure and the vectorial nature of electromagnetic fields cannot be disregarded. In these situations, optical vortices occur very rearly since all of the three field components need to be exactly zero in order to have a real singularity. Over a frequency cycle, the real part of the complex vector draws an ellipse (the polarization ellipse) and, as we proceed to detail, singularities arise as \textit{Polarization Singularities} either when light is circularly polarized or linearly polarized. Following the formalism introduced by M. R. Dennis and M. V. Berry\cite{Dennis:2001vi, Berry:2004da}, the vectors defining the major ($\mathbf{\alpha}$) and minor ($\mathbf{\beta}$) axes of the polarization ellipse and its surface normal ($\mathbf{N}$) (see Fig.\ref{polarization_ellipse}) can be expressed as:
 \begin{figure}[!tb]
\centering
 \includegraphics[width=0.5\columnwidth]{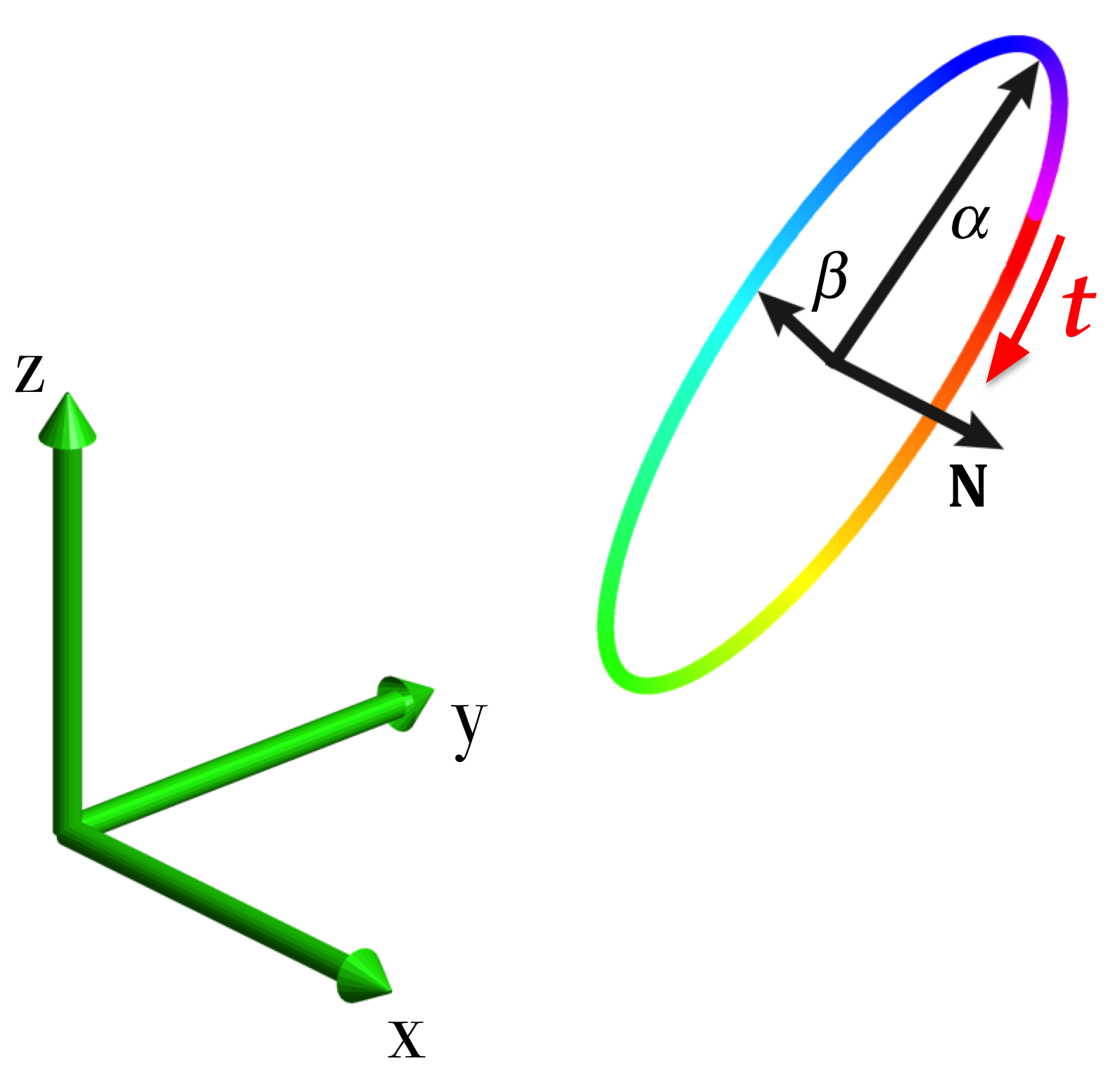}
 \caption{\label{polarization_ellipse} Polarization ellipse traced by an arbitrary electric field as a function of time ($\it{t}$). Green arrows represent the cartesian coordinates,  $\mathbf{\alpha}$ and $\mathbf{\beta}$ the major and minor axes of the polarization ellipse  and $\mathbf{N}$ its surface normal.}
 \end{figure}
 
\begin{equation}\label{alpha_eq}
\mathbf{\alpha}=\frac{1}{|\sqrt{\mathbf{E} \cdot \mathbf{E}}|}\Re\left({\mathbf{E^*}\sqrt{\mathbf{E} \cdot \mathbf{E}}}\right)
 \end{equation}
\begin{equation}\label{beta_eq}
\mathbf{\beta}=\frac{1}{|\sqrt{\mathbf{E} \cdot \mathbf{E}}|}\Im\left({\mathbf{E^*}\sqrt{\mathbf{E} \cdot \mathbf{E}}}\right)
 \end{equation}
\begin{equation}\label{N_eq}
\mathbf{N}=\Im\left({\mathbf{E}^*\times \mathbf{E}}\right)
 \end{equation}
where $\mathbf{E}=E_x\mathbf{\hat{x}}+E_y\mathbf{\hat{y}}+E_z\mathbf{\hat{z}}$ is the complex electric field vector, $\Re( \cdot)$ and $\Im( \cdot)$ represent the real and imaginary parts of a complex function and the polarization scalar $\varphi=\mathbf{E} \cdot \mathbf{E}=E_xE_x+E_yE_y+E_zE_z$. 
 
If light is linearly polarized, the polarization ellipse collapses to a line and it is not possible to define a unique vector which is normal to the polarization ellipse. In those situations, $\mathbf{|N|}=0$. In three dimensions, the spatial positions where this condition is fulfilled arrange in surfaces, named $L$ surfaces. 

On the other hand, when light is circularly polarized, the major and minor axes of polarization cannot be unambiguously determined, and both $\mathbf{\alpha}$ and $\mathbf{\beta}$ vectors become identically zero. Since $\varphi$ is a complex scalar function, its vanishing requires both its real and the imaginary parts to be zero. Thus, in $3D$ space, the points of circular polarization form lines ($C$ lines). On the other hand, as in scalar optical vortices, the phase of $\varphi$ around any zero spans all the possible values from $0$ to $2\pi l$, $l$ being the topological charge of the singularity. Around any small closed curve surrounding a C line, the major and minor axes of the polarization ellipse generate a multi twist ribbon with $l$ half twists \cite{Freund:2005tx,Freund:2005vl}. Thus, as it was recently experimentally demonstrated \cite{Bauer:2015tk}, for $l=1$, both $\mathbf{\alpha}$ and $\mathbf{\beta}$ form a M\"{o}bius strip around any small closed path arund a $C$ line. 

Since the seminal contributions by Nye \cite{Nye:1983hr,Nye:1983et,Nye:1999uv} polarization singularities in optical fields have been studied extensively for their intrinsic theoretical interest. $C$ lines and $L$ surfaces have been identified in contexts as disparate as the skylight \cite{Berry:gc}, speckle fields \cite{Flossmann:2008ua,Egorov:2008ur}, tightly focused beams \cite{Lindfors:2007bv,Schoonover:2006ks}, crystal optics \cite{Flossmann:2005uf}, photonic crystals \cite{Burresi:2009vz, Rotenberg:2015cy} and plasmonic systems \cite{deHoogh:2014db,deHoogh:2015jb}. Nevertheless, experimental applications of such topological features remained elusive until very recently. In the recent past, polarization singularities have proved to be useful in quantum information applications \cite{LeFeber:2015ul,Young:1707250}. 

In this paper, we explore the possibility of creating such polarization singularities in optical antennas illuminated by linearly polarized light. We prove that systems presenting a simultaneous isotropic electric and a magnetic polarizability ($\alpha_e$ and $\alpha_m$ respectively) are among the simplest nanostructures capable of sustaining $C$ lines and $L$ surfaces. Moreover, we prove the anomalous scattering Kerker conditions based on singular optics arguments alone. We verify our ideas by performing full field simulations on the optical response of a high index dielectric nanosphere iluminated by a linearly polarized planewave. We track the $C$ lines generated by a Si nanoparticle over all the radiation regions and unveil the M\"{o}bius strip structure of the main axis of the polarization ellipse around them. 

\section{Results and Discussion}

Dielectric nanoantennas made of materials with a high index of refraction have been intensely studied in the recent past due to their ability to sustain lossless electric and magnetic resonances in the visible and IR parts of the spectrum \cite{garcia2011strong}. On the one hand, their lossless character has proved crucial in the fabrication of metamaterials and metasurfaces not limited by ohmic losses \cite{garcia2011strong,Moitra:2013bu,Arbabi:2015fo}. On the other hand, the combination of electric and magnetic dipolar responses allows for an additional degree of freedom in the manipulation of light that has proved instrumental in surface enhanced chiral spectroscopy \cite{garcia2013surface}, and in shaping the radiation properties of optical antennas \cite{Geffrin:2012kq, Person:2013vf}. In this article, we focus in the polarization aspect of the fields scattered by such resonators illuminated by linearly polarized light. 

\begin{figure*}[!tb]
\centering
 \includegraphics[width=1\columnwidth]{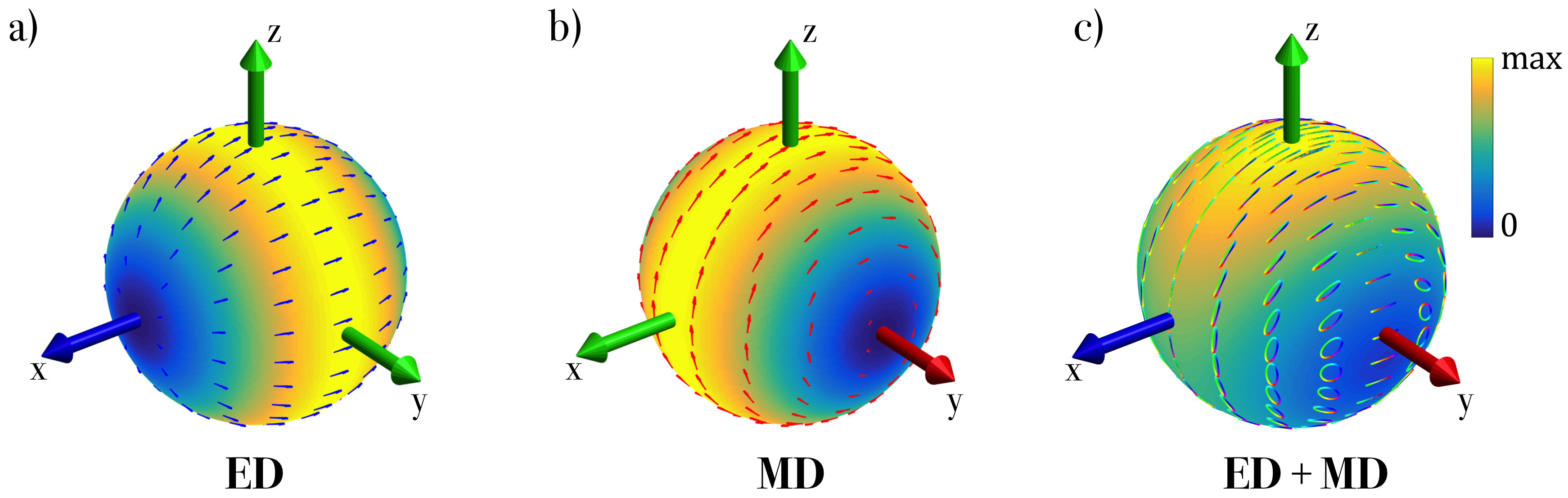}
 \caption{\label{Dipoles_scheme} Electric far-field amplitude and polarization distribution of: (a) an electric dipole, blue arrow, (b) a magnetic dipole, red arrow, and (c) a combination of an electric and a magnetic dipole. For isolated electric and magnetic dipoles (a and b) iluminated by linearly polarized light, all the scattered fields are linearly polarized. On the contrary, for a system radiating as a simultaneous set of electric and magnetic dipoles (c), interference makes the scattered field to be generally elliptically polarized. }
 \end{figure*}

Let us start by considering the far field radiation characteristics of an electric dipole. Excited by linearly polarized light ($\mathbf{E_{inc}}=E_0\mathbf{\hat{x}}$, $\mathbf{H_{inc}}=H_0\mathbf{\hat{y}}$), the induced electric dipolar moment can be expressed as $\mathbf{p}=\varepsilon_0\alpha_e\mathbf{E_{inc}}=E_0\varepsilon_0\alpha_e\mathbf{\hat{x}}$, $\varepsilon_0$ being the vacuum permittivity and $\alpha_e$ the electric dipolar polarizability 
The far field scattering of such dipolar moment is given by:
\begin{equation}\label{E_ED}
\mathbf{E_{scat}^{ED}}=\frac{k^2}{\varepsilon_0}\mathbf{G_e}\mathbf{p}=\frac{k^2}{\varepsilon_0}\left[ \left( \mathbf{n} \times \mathbf{p} \right) \times \mathbf{n} \right] \frac{e^{ikr}}{4\pi r}
 \end{equation}
where $\mathbf{G_e}$ is the electric Green's tensor in the far field approximation and k is the wavenumber of light in vacuum. $r = |\mathbf{r} - \mathbf{r_0}|$ is the distance between the position of the dipole ($\mathbf{r_0}$) and the observation point  $\mathbf{r}$, and $\mathbf{n}=n_x\mathbf{\hat{x}}+n_y\mathbf{\hat{y}}+n_z\mathbf{\hat{z}}$ is the unit vector in the direction of $\mathbf{r} - \mathbf{r_0}$. Figure \ref{Dipoles_scheme}a plots the far field electric field amplitude ($|\mathbf{E_{scat}^{ED}}|$) with superimposed electric field polarization vectors. As expected, the fields scattered by a linearly polarized electric dipole are linearly polarized in all spatial directions. 
 
 The situation is similar for an isotropic magnetic dipole polarizability $\alpha_m$ excited by linearly polarized light. In such a system, the induced magnetic dipolar moment can be expressed as  $\mathbf{m}=\alpha_m\mathbf{H_{inc}}=H_0\alpha_m\mathbf{\hat{y}}$, and the scattered fields in the far field can be calculated from:
 \begin{equation}\label{E_MD}
\mathbf{E_{scat}^{MD}}=iZ_0k^2\mathbf{G_m}\mathbf{m}=-Z_0k^2\left( \mathbf{n} \times \mathbf{m} \right) \frac{e^{ikr}}{4\pi r}
 \end{equation}
where $Z_0$ in the impedance of free space and $\mathbf{G_m}$ is the magnetic Green's tensor in the far field approximation. Figure \ref{Dipoles_scheme}b presents the far field electric field amplitude ($|\mathbf{E_{scat}^{MD}}|$) with superimposed electric field polarization vectors. In this case, the electric field vectors circulate around the $\mathbf{\hat{y}}$ polarized magnetic dipole, but the polarization is also linear in the entire solid sphere.

Nevertheless, this situation changes when considering the scattering coming from a particle with simultaneous isotropic electric and magnetic polarizabilities excited by linearly polarized light. In this case, the scattered fields are a superposition of the fields radiated by both the electric and magnetic dipole:
 \begin{equation}\label{E_EDMD}
\mathbf{E_{scat}^{ED+MD}}=\frac{k^2}{\varepsilon_0}\mathbf{G_e}\mathbf{p} + iZ_0k^2\mathbf{G_m}\mathbf{m}.
 \end{equation}

$\mathbf{E_{scat}^{ED}}$ and $\mathbf{E_{scat}^{MD}}$ are not necessarily in phase and they are not parallel. Thus, the polarization state in these systems (Figure \ref{Dipoles_scheme}c) becomes spatially inhomogeneous. Since the scattered fields become generally elliptically polarized, it is logical to think that these kind of systems may sustain polarization singularities. 
 
We will first search for $L$ surfaces. Calculating $|\mathbf{N}|$ by substituting Eq.\ref{E_EDMD} into Eq.\ref{N_eq}, one obtains:
 \begin{equation}\label{N_an}
|\mathbf{N}|=\Im {\left(\alpha_e \alpha_m^*\right)}n_x n_y.
 \end{equation}
Thus, both the $\hat{xz}$ plane and the $\hat{yz}$ plane are $L$ surfaces ($|\mathbf{N}|=0$) in the radiation region. 

The analytical determination of $C$ lines is more intricate. Deriving the polarization scalar $\varphi$ from Eq.\ref{E_EDMD} yields:
 \begin{equation}\label{phi_an}
\varphi=\alpha_m^2 n_x^2 +\alpha_e^2 + (\alpha_e^2+\alpha_m^2) n_z^2 + 2 \alpha_e\alpha_m n_z.
 \end{equation}

Consequently, light will be circularly polarized if
\begin{equation}\label{phi_re}
\Re{\left(\alpha_m^2 n_x^2 +\alpha_e^2 n_y^2+ (\alpha_e^2+\alpha_m^2) n_z^2 + 2 \alpha_e\alpha_m n_z\right)}=0
\end{equation}
\begin{equation}\label{phi_im}
\Im{\left(\alpha_m^2 n_x^2 +\alpha_e^2 n_y^2+ (\alpha_e^2+\alpha_m^2) n_z^2 + 2 \alpha_e\alpha_m n_z\right)}=0
 \end{equation}.
 
 Since $\mathbf{n}$ is a unit vector, a third condition also holds, 
 \begin{equation}\label{phi_n}
n_x^2+n_y^2+n_z^2=1
 \end{equation}

Note that none of these equations do depend on $r$. Thus, for some particular $\alpha_e$ and $\alpha_m$, $n_x$, $n_y$ and $n_z$ are the only unknowns in this set of 3 equations and their solutions specify the directions on which the $C$ lines propagate on the far field. It is interesting to note that geometrically, Eq.\ref{phi_re} and Eq.\ref{phi_im} describe two hyperbolic functions, and that Eq. \ref{phi_n} describes a unitary sphere. So, for particular values of the complex electric and magnetic polarizability, the direction of the $C$ lines will be determined by the intersection of these three surfaces. An analytic solution to this equations does exist, but unfortunately, it is too cumbersome to provide any intuitive interpretation. Nevertheless some interesting conclusions can be extracted under some particular assumptions. 

If $\alpha_e=\alpha_m$, it can be easily proved that equations \ref{phi_re}-\ref{phi_n} determine that a single $C$ line exists in the direction of $\mathbf{n}=-\hat{\mathbf{z}}$. Nonetheless, according to Eq.\ref{N_an}, the $\hat{\mathbf{z}}$ axis belongs to a $L$ surface. So, according to this formalism, light should be both linearly and circularly polarized in the back-scattering trajectory. This is only feasible if no fields are radiated in this direction. This phenomenon is the well known Kerker condition for zero back-scattering \cite{Kerker:1983kb, Geffrin:2012kq, Person:2013vf,ZambranaPuyalto:2013cl}, but, to the best of our knowledge, it has never been proved before based on singular optics arguments. For $\alpha_e=-\alpha_m$, the second Kerker condition for zero forward-scattering can also be easily derived following the same procedure. 

\begin{figure}[!tb]
\centering
 \includegraphics[width=0.8\columnwidth]{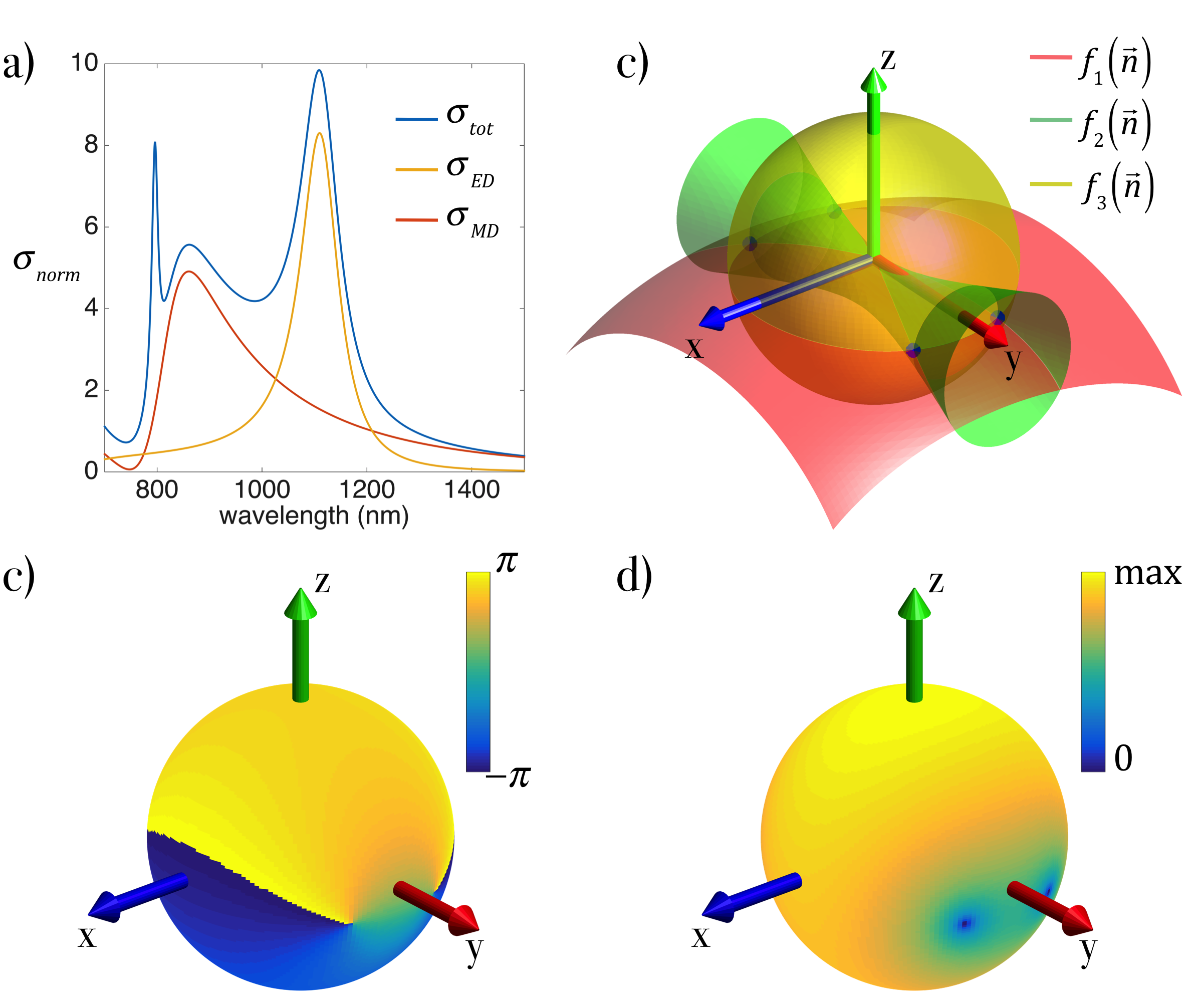}
 \caption{\label{FF_Clines} (a) Extinction cross section of a Si nanosphere (150 $nm$ radius) illuminated by linearly polarized light. Blue line represents the geometrically normalized total extinction cross section. The yellow and red lines are the magnetic and electric dipolar contributions to the total extinction cross section. (b) Geometrical solution to equations \ref{phi_re} - \ref{phi_n} for $\lambda=1110$ $nm$. Eq.\ref{phi_re} (red surface) and Eq.\ref{phi_im} (green surface) describe two hyperbolic functions, while Eq. \ref{phi_n} (yellow surface) describes the unitary sphere. (c) Phase and (d) amplitude of $\varphi$ as a function of position, for the 150 $nm$ Si sphere illuminated by a $1110$ $nm$ linearly polarized plane wave, calculated in the far-field, 20 $\mu$m away from the origin. Blue and red arrows indicate the direction of the induced electric and magnetic dipoles respectively. As expected, zeros in (d) correspond to phase singularities in (c), indicating the presence of polarization singularities in those directions. }. 
 \end{figure}

To verify the existence of $C$ lines in a realistic system, 
we consider the optical response of a 150 $nm$ silicon sphere illuminated by linearly polarized light. Figure \ref{FF_Clines}a shows the geometrically normalized extinction cross section of this system (blue line) calculated using Mie theory. 
 Yellow and red lines plot the contribution of the dipolar electric and magnetic terms in the Mie series to the total extinction spectrum. As expected, the lowest energy peak at 1110 $nm$ corresponds to the first order $b$ term in the Mie series ($b_1$) . This resonance is of magnetic dipolar character and it can be associated to a magnetic polarizability through \cite{garcia2011strong} $\alpha_m=\frac{i}{\mu_0}\left(\frac{k^3}{6\pi}\right)^{-1}b_1$. On the contrary, the peak at 860 $nm$ is the lowest order electric resonance and its dipolar polarizability can be inferred from the $a_1$ term in the Mie series through $\alpha_e=i\varepsilon_0\left(\frac{k^3}{6\pi}\right)^{-1}a_1$. Figure \ref{FF_Clines}b plots the surfaces defined by equations \ref{phi_re}, \ref{phi_im} and \ref{phi_n} for $\lambda=1110nm$. Four intersection points can be identified (blue dots). These points indicate the directions on which $C$ lines generated by the silicon nanoparticle will propagate through the far field. 
 
To verify our analytical predictions,  we solve exactly Maxwell's equations using the Boundary Element Method \cite{GarciadeAbajo:2002kt, GarciadeAbajo:1998hb, Hohenester:2012vj} for such a system. In particular, we calculate $\varphi$ at a distance 20 $\mu m$ away from the Si nanoantenna. Figure \ref{FF_Clines}c and Figure \ref{FF_Clines}d plots the phase of $\varphi$ and its absolute value. It easy to see that, at the predicted positions, $|\varphi|$ presents exact zeros while the phase becomes singular. Along these lines, the scattered fields are exactly circularly polarized, their handedness being determined by the sign of $\mathbf{n}\cdot \mathbf{N}$ which is positive for right handed fields and negative left handed fields [ref Dennis thesis]. 



\begin{figure*}[!tb]
\centering
 \includegraphics[width=1\columnwidth]{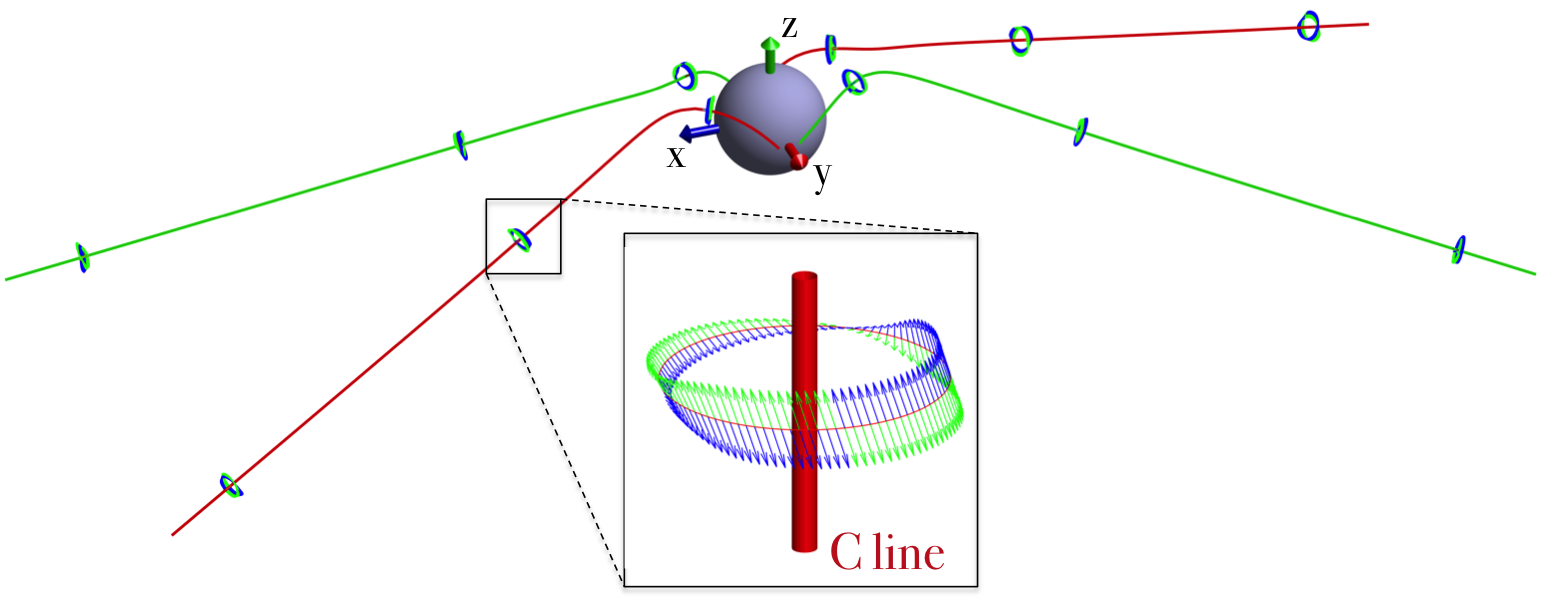}
 \caption{\label{Clines_3D} $C$ lines around the Si nanoparticle for $\lambda=1110nm$ in all the radiation regions. The silicon nanoparticle is represented by the blue sphere. Green and red lines are the left and right handed $C$ lines respectively. The main axis of the polarization ellipse forms a single twist M\"{o}bius strip in any curve enclosing a $C$ line. The inset is a zoom of this polarization structure around one particular $C$ line. }
 \end{figure*}

Finally, we calculate the polarization structure on a 3D volume encompassing all the radiation regions around the 150 $nm$ Si nanoparticle excited by a 1100 $nm$ linearly polarized plane wave. We track the $C$ lines from the near-field to the far-field by the procedure described in the Methods section. Figure \ref{Clines_3D}, compiles these results. The blue sphere represents the Si nanoparticle while the red and green lines emerging from the sphere are the left and right handed $C$ lines respectively. These calculations reveal that even though in the far field the $C$ lines evolve radially as straight lines, in the near-field they curve in a non trivial manner. Moreover, according to Freunds's predictions \cite{Freund:2005tx,Freund:2005vl}, taking a closed loop around any point on a $C$ line, any of the axes of the polarization ellipse should form a multi-twist strip \cite{Bauer:2015tk}, with the number of twists being the equal to the topological charge of the $C$ line. In this particular case, since $|l|=\frac{1}{2}$ for all of the C lines, a M\"{o}bius strip should be formed around any loop around any of the $C$ lines. We verify this polarization structure through the calculation of the major axis of the polarization ellipse on several circles encompassing the $C$ lines along their trajectory. The inset in figure \ref{Clines_3D}, is a zoom of one of these calculations. For the purpose of displaying the M\"{o}bius strip clearer, one half of the major axis of the polarization ellipse is depicted as a green arrow and the other half in blue. The single twist in the ribbon is evident and consistent all along every $C$ line.
 
\section*{Conclusions}
 
In conclusion, we have analytically and numerically proved that high index nanoparticles support polarization singularities that propagate from the near-field to the far-field. In particular, $C$ lines arise from the interference of the fields scattered by electric and magnetic isotropic dipolar excitations. The study of polarization singularities is a rapidly growing field of research, on one hand due to the intrinsic theoretical interest of the topology of light fields and on the other hand because of the upcoming applications on quantum information systems. This theoretical contribution proves that combinations of electric and magnetic dipoles supported by high index nanoparticles create such topologically protected features and may facilitate new experimental platforms to study polarization singularities.

\section*{Methods}
\subsubsection*{Identification and tracking of $C$ lines around optical antennas}

In order to identify and track the $C$ lines emerging from an optical antenna, we first calculate the vectorial electric fields on a meshed surface covering the nanoparticle at a short distance form its periphery  ($\mathbf{E_0(r)}$). In particular, to compute the results presented in Fig. \ref{Clines_3D}, we choose to use the Metallic Nanoparticle Boundary Element Method MATLAB toolbox \cite{Hohenester:2012vj}. Starting from $\mathbf{E_0(r)}$, the polarization scalar can be readily computed as:
\begin{equation}
\varphi_0(\mathbf{r})=u(\mathbf{r})+iv(\mathbf{r})=\mathbf{E_0(r)} \cdot \mathbf{E_0(r)}=E_x(\mathbf{r})E_x(\mathbf{r})+E_y(\mathbf{r})E_y(\mathbf{r})+E_z(\mathbf{r})E_z(\mathbf{r}).
 \end{equation}

 The objective is to find the exact zeros of $\varphi$ on the surface surrounding the nanoparticle. In order to do so, we first find the local minima of $|\varphi|$ on the calculated mesh, and then run a minimization routine (fminsearch.m in Matlab) on $|\varphi(\mathbf{r})|$ using the identified minima as seeds for the minimization routine. After verifying that these minima are actual zeros of $\varphi(\mathbf{r})$, we take them as the starting points of the $C$ lines ($\mathbf{r_0}$). The direction of the $C$ line can be calculated though the following expression: \cite{Dennis:2001vi}
 
 \begin{equation}
\mathbf{d(r_0)}=\frac{1}{2}\nabla \varphi^*(\mathbf{r_0}) \times \nabla \varphi(\mathbf{r_0}) =\nabla u(\mathbf{r_0})  \times \nabla v(\mathbf{r_0})  
 \end{equation}
 
 Having calculated the origin of the $C$ line and its direction, it is easy to calculate the seed of the next point on the $C$ line as 
 \begin{equation}
 \mathbf{r_1^s}=\mathbf{r_0}+l\mathbf{d(r_0)}, 
 \end{equation}
 with 
 \begin{equation}
 l=l_0\frac{\mathbf{d(r_0)} \cdot\mathbf{\hat{r}}}{|\mathbf{d(r_0)} \cdot\mathbf{\hat{r}}|}
 \end{equation}
 being the target distance to the next point on the line, $l_0$, corrected for the fact that $\mathbf{d}$ may point towards the nanostructure. Minimizing $\varphi$ around this new position, it is easy to calculate the exact position of the next point on the $C$ line. One can compute the entire trajectory of the $C$ line by following this procedure recursively. 

\begin{acknowledgement}
We thank J. J Saenz for insightful discussions and XX for the careful reading of the manuscript.  A. G.-E. received funding from the Fellows Gipuzkoa fellowship of the Gipuzkoako Foru Aldundia through FEDER "Una Manera de hacer Europa"
\end{acknowledgement}

\begin{suppinfo}
\end{suppinfo}


\bibliography{singular_optics}

\end{document}